\documentclass[final,3p,times,twocolumn]{elsarticle}

\usepackage{epsfig,bm}
\usepackage{graphics}
\usepackage{amsmath}
\usepackage{mathrsfs}
\usepackage[normalem]{ulem}  
\usepackage[dvips]{color} 

\renewcommand\sout{\bgroup \color{red} \ULdepth=-.5ex \ULset}

\biboptions{sort&compress}

\begin{document}

\begin{frontmatter}


\title{Heavy quark correlations and the effective volume  for quarkonia production}

\author[ic]{Yunpeng Liu}
\author[ic,ph]{Che Ming Ko}
\author[ic,ph]{Feng Li}

\address[ic]{Cyclotron Institute, Texas A$\&$M University, College Station, Texas 77843, USA}
\address[ph]{Department of Physics and Astronomy, Texas A$\&$M University, College Station, Texas 77843, USA}%
\begin{abstract}
Using the Boltzmann transport approach, we study the effective volume of a correlated heavy quark pair in a partonic medium based on their collision rate.  We find that the effective volume is finite and depends sensitively on the momentum of the heavy quark and the temperature of the medium. Generally,  it increases linearly with time $t$ at the very beginning and the increase then becomes slower due to multiple scattering, and finally it increases linearly with respect to $t^{3/2}$.   We further find that the colliding heavy quark pair has an effective temperature  similar to that of the medium even though their initial transverse momentum spectra are far from thermal equilibrium. 
\end{abstract}

\begin{keyword}
heavy ion collision, quark-gluon plasma, heavy flavor
\PACS{25.75.-q, 24.85.+p}
\end{keyword}
\end{frontmatter}



\section{Introduction}

 The conservation of Abelian charges plays an important role for particle production in heavy ion collisions.  One example is the so-called canonical suppression of strange particle production  as a result of the conservation of strangeness~\cite{Hagedorn, Hagedorn:1984uy, Cleymans:1996yg, Hamieh:2000tk, Ko:2000vp,Fochler:2006et}, which requires that a hadron consisting of a strange quark is always produced together with another hadron consisting of an antistrange quark. Because of their correlated production, the two hadrons are initially close in space and their subsequent annihilation probability thus differs from the case when they are initially randomly distributed in space.  As shown in Ref.~\cite{Pal:2001nk}, this has a significant effect on their chemical equilibration time in heavy ion collisions.   A similar effect is expected for the production of hadrons consisting of heavy charm or bottom quarks~\cite{Gorenstein:2000ck, Andronic:2006ky} due to the conservation of charm or bottom.  In particular, since the heavy quark and antiquark pair  are produced at the same location and  their momenta are also highly correlated according to the leading order QCD,  such correlations , especially when there is at most one pair of heavy quarks in an event, would then affect their subsequent collision rate and thus the rate for quarkonia production~\cite{Grandchamp:2003uw}.
\par
The dynamics of heavy quarks in a medium has been widely studied using the Langevin or the Fokker-Planck equation~\cite{Svetitsky:1987gq, vanHees:2004gq, Rapp:2009my, He:2014cla, Cao:2011et, Young:2011ug, Alberico:2011zy}.  These studies have shown that the observed suppression of heavy quarks at high momentum implies strong interactions between heavy quarks and partons in the medium.  However, it was recently pointed out that the Boltzmann equation gives a more accurate description of the dynamics of heavy quarks in a medium~\cite{Das:2012ck}. In the present paper, we adopt the Boltzmann equation to study the motion of a pair of heavy quarks that are initially correlated in both the coordinate and the momentum space, as produced in high energy collisions, to study the likelihood they would scatter again in terms of an effective volume. 

\section{The model}

Given two particle species $A$ and $B$ in a volume $V$ at time $t$, their collision number $\Delta N_{AB}$ in a small time step $\Delta t$\ can be expressed as
\begin{eqnarray}
   \Delta N_{AB}&=&\Delta t\int_V d{\bf x}\int \frac{ d{\bf p}_{A}d{\bf p}_{B}}{(2\pi)^3(2\pi)^3}f_{A}({\bf x}, {\bf p}_{A}, t)\nonumber\\
   &&\cdot f_{B}({\bf x}, {\bf p}_{B}, t)v_{AB}\sigma_{AB}.
   \label{eq_dN_1}
\end{eqnarray}
In the above, the distribution functions $f_{A}$ and $f_{B}$  are functions of the position ${\bf x}$ and momenta ${\bf p}_{A}$ and ${\bf p}_{B}$,  while $v_{AB}$\ and $\sigma_{AB}$  are their relative velocity and total scattering cross section.
\par
For heavy quark $Q$ and antiquark $\bar{Q}$, if we assume that the distribution functions  $f_Q=f_{\bar{Q}}$ are uniform in space and have a Boltzmann distribution in momentum with a temperature $T$, then the collision number is simply given by
\begin{eqnarray}
   \Delta N_{Q\bar{Q}}^{\rm th}&=&\frac{TV\Delta t}{4\pi^4}\frac{N_QN_{\bar{Q}}}{N_{Q_0}N_{{\bar Q}_0}}\int_{\sqrt{s_0}}^{\infty} d\sqrt{s}\ sp^2 \sigma_{Q\bar{Q}}(\sqrt{s})\nonumber\\
   &&\cdot K_1(\sqrt{s}/T),
  \label{eq_ncoll_s}
\end{eqnarray}
where $\sqrt{s_0}=m_Q+m_{\bar{Q}}$ is the minimum center of mass energy $\sqrt{s}$ of the scattering pair,  $p= \sqrt{[(s-m_Q^2-m_{\bar{Q}}^2)^2-4m_Q^2m_{\bar{Q}}^2]/(4s)}$ is their momentum in their center of mass frame, and 
$K_1$ is the modified Bessel function of the second kind. The number of  $Q$ in a volume $V$ and its thermally 
equilibrated number with vanishing chemical potential are denoted by $N_Q$ and  $N_{Q_0}= m_Q^2VTK_2(m_Q/T)/(2\pi^2)$,  respectively, while those of  $\bar{Q}$ are, respectively,  $N_{\bar Q}$ and  $N_{{\bar Q}_0}$.  In the case of a constant cross section  $\sigma_{Q\bar{Q}}$  and equal masses $ m_Q=m_{\bar{Q}}=m$,  the above expression can be further expressed as
\begin{eqnarray}
   \Delta N_{Q\bar{Q}}^{\rm th} &=& \frac{N_{Q}N_{\bar{Q}}\sigma_{Q\bar{Q}} \Delta t}{V}g(m^*)
   \label{eq_th}
\end{eqnarray}
with $g(z)\equiv 4K_3(2z)/[zK_2^2(z)]$\ and $m^*=m/T$.
\par
On the other hand, if only one pair of heavy quarks is produced in a heavy ion collision,  they will initially be at the same position and have essentially same momentum but with opposite direction.  As they propagate in the medium, they will gradually move apart and also become thermalized. Because of the large size of the medium and the relatively long relaxation time for heavy quarks and antiquarks in the medium, they are not likely to trace the whole volume of the fireball and to achieve kinetic equilibrium immediately.  The deviation of the heavy quark distribution function from the uniform and thermal distribution used in obtaining Eq.~(\ref{eq_th}) for rare heavy quark events in relativistic heavy ion collisions (e.g. charm at SPS energy and bottom at RHIC energy) can be characterized by an effective volume,
\begin{eqnarray}\label{volume}
   V_{\textrm{eff}}&\equiv & \lim_{\substack{\Delta t\rightarrow 0, \sigma_{Q\bar Q}\rightarrow 0,\\ N_{Q}, N_{\bar Q}\rightarrow \infty}}\frac{N_{Q}N_{\bar Q}\sigma_{Q\bar Q} \Delta t}{\Delta N_{Q\bar Q}}g(m^*),
   \label{eq_V_eff}
\end{eqnarray}
which only depends on the distribution of the heavy quark pairs in the phase space and the temperature of the medium. It approaches the volume of the medium when the heavy quarks are thermalized.
\par
The nonequilibrium dynamics of heavy quarks can be described by the Boltzmann equation for their phase space distribution function $f_Q({\bf x},{\bf p},t),$
\begin{eqnarray}
   \partial_t f_Q({\bf x}, {\bf p}, t) + {\bf v}_Q\cdot\nabla f_Q({\bf x}, {\bf p}, t) &=& C[f_Q],
\end{eqnarray}
where  ${\bf v}_Q$\  is the velocity of the heavy quark, and $C[f_Q]$\ is the collision term. For a given initial position and momentum of a heavy quark, this equation can be solved using the 
heavy quark collision rate with thermal partons in the medium, given by
\begin{eqnarray}
   R&=& \frac{mT^3m_{g}^{*2}N_g}{2\pi^2E}\int_0^{\infty} d y \ e^{-m_{g}^*\cosh y\cosh y_r}\sinh^2y\nonumber\\
  &&\cdot\sinh(m_{g}^*\sinh y_r\sinh y)\sigma_{gQ}(y),
\end{eqnarray}
which is obtained directly from Eq.~(\ref{eq_dN_1})  by using the Boltzmann distribution 
for the thermal partons. In the above, $\sigma_{gQ}(y)$ is the cross section for scattering between the heavy quark and a parton of rapidity $y$  in the heavy quark frame; $N_g$ is the degeneracy of the parton, which is taken to be 16 if we include only gluons and neglect quarks as their scattering cross sections with heavy quarks are small~\cite{Rapp:2009my}; $y_r=\textrm{acosh}(p\cdot u/m)$ is the rapidity of the heavy quark relative to the medium with four-velocity  $u^{\mu}$;  $m_g^*=m_g/T$  where $m_g$  is the mass of gluons; and $E$ is the energy of the heavy quark. For a constant cross section and massless partons,  the heavy quark collision rate in the rest frame of the medium can be simplified to $R= N_gT^3\sigma_{gQ}/\pi^2$,  resulting in a mean free time between collisions given by the inverse $\tau=1/R$ . As the heavy quark traverses through the medium, the probability for it to collide with a parton during a small time step $dt\ll \tau$ is then $dt/\tau$. A collision occurs if a random number generated between 0 and 1 is smaller than $dt/\tau$.  With the parton momentum randomly selected according to the thermal distribution,  the momentum of the heavy quark after the collision can be determined from the energy and momentum conservations once its direction is obtained from the differential cross section.  For simplicity, we take the cross section to be isotropic with the magnitude $\sigma_{gQ}=4$\ mb, which is the approximate value expected from the pQCD for a high energy heavy quark scattering with a thermal parton~\cite{Rapp:2009my}. The treatment to $\bar{Q}$\ is identical to $Q$, except that the directions of their initial momenta are opposite.
\par
For the scattering between the heavy quark and antiquark pair, it is treated by comparing their impact parameter with the scattering cross section and is described in detail in \ref{ap_b}. Although only one pair of heavy quarks is initially produced at the same position with opposite momentum in a heavy ion collision event, we can study their mean dynamics in the hot medium by following independently the motions of many similar pairs of heavy quarks and calculate the average number of heavy quark scatterings as  in Eq.~(\ref{eq_V_eff}). 

\section{Results and discussions}

\subsection{ Heavy quarks in a closed thermal system}

\begin{figure}[!hbt]
    \centering
    \includegraphics[width=0.45\textwidth]{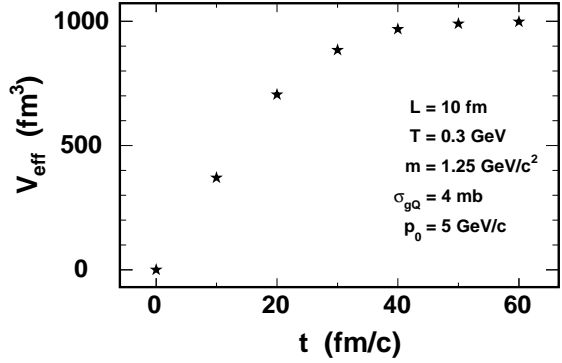}
    \caption{Time evolution of the effective volume $V_{\textrm{eff}}$ of a pair of heavy quarks with mass $m$ and back-to-back momentum $p_0$ in a periodic cubic box of length $L$ on each side, which consists of a gluonic matter at temperature $T$, and undergo scattering with gluons with the cross section $\sigma_{gQ}$.}
    \label{fig_box}
\end{figure}

To illustrate the method used in our study,  we first consider the collision dynamics of a pair of heavy 
quarks in a periodic cubic box of length  $L=10$ fm on each side. The heavy quarks are initially located at same position and have opposite momentum of $5$ GeV$/c$.  With the temperature of the medium taken to be $0.3$\ GeV, the time evolution of the effective volume $V_{\textrm{eff}}$ of the heavy quark pair calculated according to Eq.~(\ref{volume}) is shown in Fig.~\ref{fig_box}.  As expected, the effective volume $V_{\textrm{eff}}$ approaches the volume of the medium as time $t$ approaches infinity, and the time scale for the heavy quarks to spread uniformly in the box is about $30$ fm$/c$.

\begin{figure}[!hbt]
    \centering
    \includegraphics[width=0.45\textwidth]{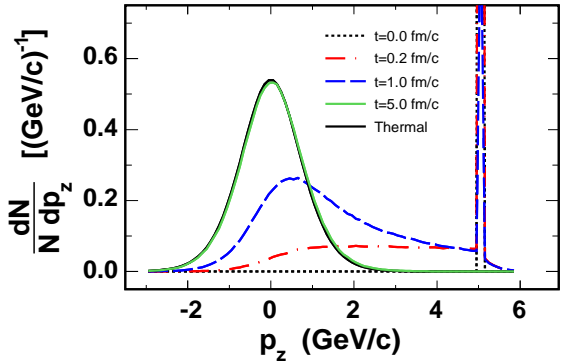}
    \caption{(Color on line) Longitudinal momentum $p_z$ distribution of heavy quarks with initial momentum $p_{0}=5$\ GeV$/c$ along the $z$ direction at different times in a finite gluonic matter of temperature $T=0.3$ GeV and the thermal $p_z$ distribution calculated from Eq.~(\ref{eq_pzdis_th}) }
    \label{fig_pzdis}
\end{figure}
\par

To see how heavy quarks approach thermal equilibrium, we show in Fig.~\ref{fig_pzdis} their  $p_z$ distribution at different times. It is seen that the initial $\delta$-like peak at $p_z=5$ GeV$/c$ initially moves down in $p_z$ and gradually approaches a thermal distribution.  During early times, the distribution deviates significantly from a Gaussian function, in contrast with the prediction based on the Langevin approach because of the comparable average parton kinetic energy and heavy quark mass, which is beyond the region where the Langevin approach is applicable~\cite{Das:2013kea} .  Since heavy quarks have a mean free time $\tau=0.44$ fm$/c$, their  $p_z$\ distribution essentially becomes the thermal distribution
\begin{eqnarray}
   f_{P_z}(p_z)=\frac{dN}{N\ dp_z}&=& \frac{(m_L^*+1)e^{-m_L^*}}{2Tm^{*2}K_2(m^*)}
   \label{eq_pzdis_th}
\end{eqnarray}
with $m_L^*\equiv \sqrt{m^2+p_z^2}/T$ at $t=5$\ fm$/c\ \gg\tau$.

\subsection{ Heavy quarks in an open thermal system}

\begin{figure}[!hbt]
    \centering
    \includegraphics[width=0.45\textwidth]{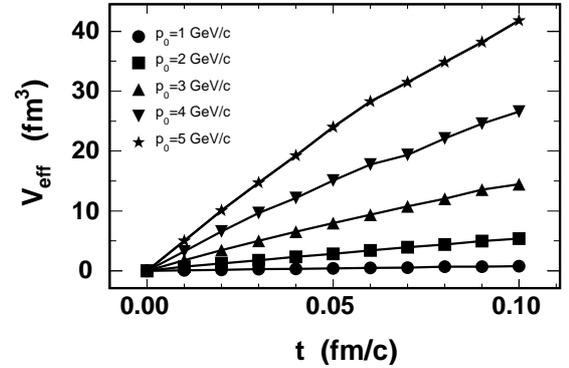}
    \caption{Time evolution of $V_{\textrm{eff}}$ at early times $t\ll\tau$ for different initial heavy quark momentum $p_0$ in an infinite medium of temperature $T=0.3$ GeV.}
    \label{fig_Veff_short}
\end{figure}

\begin{figure}[!hbt]
    \centering
    \includegraphics[width=0.45\textwidth]{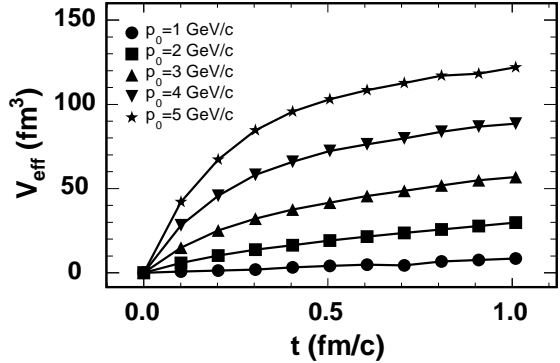}
    \caption{Same as Fig.~\ref{fig_Veff_short} for intermediate times $t\sim\tau$.}  
    \label{fig_Veff_mid}
\end{figure}

\begin{figure}[!hbt]
    \centering
    \includegraphics[width=0.45\textwidth]{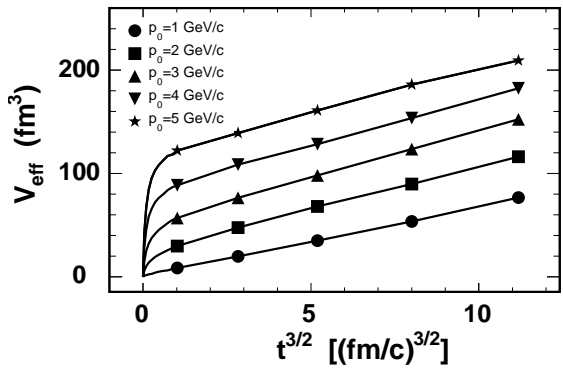}
    \caption{Same as Fig.~\ref{fig_Veff_short} for long times $t\gg\tau$ but with the horizontal axis changed from $t$\ to $t^{3/2}$.}
    \label{fig_Veff_long}
\end{figure}

In this subsection, we consider the case that the heavy quarks move in a medium of infinite volume, using the same heavy quark mass $m=1.25$ GeV$/c^2$, heavy quark-gluon scattering cross section $\sigma_{gQ}=4$ mb, and the medium temperature $T=0.3$ GeV (except in Section \ref{sss_T_depen})  as in the previous subsection.   Results on the time evolution of $V_{\textrm{eff}}$  are shown in Figs.~\ref{fig_Veff_short}, \ref{fig_Veff_mid}, and \ref{fig_Veff_long} for the three time stages $t\ll \tau$, $t\sim \tau$, and $t\gg \tau$ relative to the mean free time $\tau=0.44$ fm$/c$. In all three cases, the effective volume $V_{\textrm{eff}}$ depends sensitively on the initial momentum with the larger one resulting in a larger effective volume.

As shown in Fig.~\ref{fig_Veff_short}, the effective volume $V_{\textrm{eff}}$ at earlier times $t\ll\tau$ is proportional to time, which is true in general, and the proof is given in \ref{ap_veff}.  

When the time $t$  becomes comparable to the mean free time $\tau$, heavy quarks are more likely to turn around and collide with each 
other after undergoing successive collisions. As a result,  the increase of $V_{\textrm{eff}}$ with time becomes slower as shown in Fig.~\ref{fig_Veff_mid}. 

For time $t$ much larger than the mean free time $\tau$,  when heavy quarks have collided many times with partons in the medium, their behavior can be described by random walks.  In this picture, the distance traveled by a heavy quark is proportional to $t^{1/2}$,  and the effective volume for heavy quarks to collide is thus proportional to $t^{3/2}$ as shown in Fig.~\ref{fig_Veff_long}. Although the magnitude of the effective volume increases with heavy quark initial momentum, as it depends on its nonequilibium dynamics, the coefficient of the proportionality or the slope of the lines in the figure is independent of the initial momentum, since it only depends on their thermal motions. 

 Since the initially produced 
charm pair are  distributed in a  certain volume due to their quantum nature,  our classical calculation based on 
the Boltzmann equation is  valid only  after certain time $t_0$ 
 when the wave packets of the heavy quark and  antiquark 
are separated.\footnote{A detailed discussion  on the value of $t_0$\ requires the  spatial distribution of the charm quark when it is produced, which is beyond the scope of study in this paper.} The total number of $Q\textrm{-}\bar{Q}$ collisions is then given by 
\begin{eqnarray}
   N_{Q\bar Q}&=&\int_{t_0}^{\infty} dt\ \frac{\sigma_{Q\bar{Q}} }{V_{\textrm{eff}}(t)} g(m^*).
   \label{eq_coll}
\end{eqnarray}
Because of the long time behavior of the heavy quark effective volume, the integral in Eq.~(\ref{eq_coll}) converges at $t=\infty$. Therefore, the heavy quark pair hardly have the chance to collide with each other long after they are produced even if the lifetime of the medium is infinitely long.\footnote{The space dimension $d=3$ is important here. If the dimension is $d=1$\ or $d=2$, the heavy quark pair will collide with each other at a certain time since the integral in Eq.~(\ref{eq_coll}) diverges at $t=\infty$, and the equilibrium between the heavy quark pair and quarkonia can be established even if the system is infinitely large.} 

\par

\subsubsection{Temperature dependence of the effective volume}\label{sss_T_depen}

\begin{figure}[!hbt]
    \centering
    \includegraphics[width=0.45\textwidth]{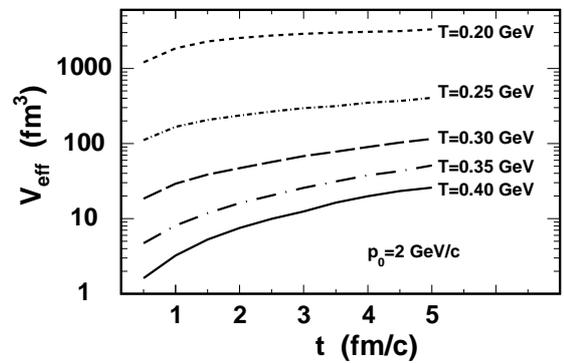}
    \caption{Time evolution of the effective volume $V_{\textrm{eff}}$  of heavy quarks with initial charm momentum $p_{0}=2$ GeV$/c$ for different medium temperatures.}
    \label{fig_T_depen}
\end{figure}

Since both the parton density and the parton energy depend on the temperature of the medium, the heavy quark effective volume also depends on the temperature of the medium, and this is shown in Fig.~\ref{fig_T_depen}.  The effective volume is seen to depend sensitively on temperature.  For the temperature $T=0.2$\ GeV, the effective volume already exceeds 1000 fm$^3$\ at $t=0.5$\ fm$/c$ due to the strong back to back correlation, while for $T=0.4$\ GeV, the effective volume is less than $10$\ fm$^3$\ at $t=2$\ fm$/c$ as a result of faster thermalization of the heavy quarks. As discussed in the previous subsection, the  $V_{\textrm{eff}}$\ increases with time monotonously for all temperatures. 
 
\subsubsection{Center of mass frame energy distribution} 
 
\begin{figure}[!hbt]
    \centering
    \includegraphics[width=0.45\textwidth]{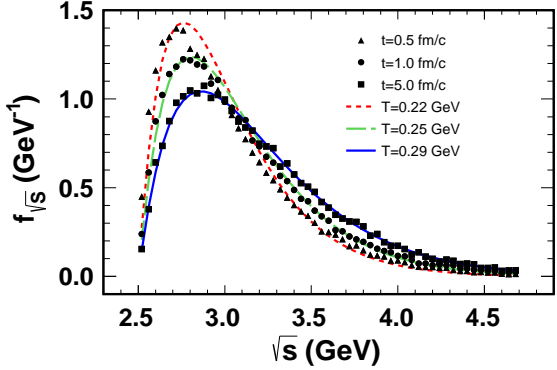}
    \caption{Center of mass frame energy $\sqrt{s}$\ distribution of colliding heavy quark pairs with initial back-to-back momentum of $5$ GeV$/c$ in a medium of temperature $T=0.3$ GeV at different times (shown by symbols).  Lines are collision rates from calculations based on thermalized heavy quarks.}
    \label{fig_ss_dis}
 \end{figure}

Since a heavy quark pair is produced from hard collisions of nucleons at high energies, their initial momenta are large and opposite in direction. As they diffuse in a medium and collide with thermal partons, their center of mass energy is a quantity of interest. Shown in Fig.~\ref{fig_ss_dis} by symbols are the distribution $f_{\sqrt{s}}$ at different times for a pair of heavy quarks with an initial back-to-back momentum of $5$ GeV$/c$ in a medium of temperature $T=0.3$ GeV. It is seen that the peak of the distribution increases to larger center of mass energy $\sqrt{s}$ with increasing time. Also shown in the figure by lines is the center of mass energy distribution of heavy quark pairs that have a thermal distribution.  
 According to Eqs.~(\ref{eq_ncoll_s}) and (\ref{eq_th}), the latter is given by
\begin{eqnarray}
   f_{\sqrt{S}}^{\textrm{th}}(\sqrt{s}, T)&\equiv&\frac{1}{ N_{Q\bar{Q}}^{\textrm{th}} }\frac{d N_{Q\bar{Q}}^{\textrm{th}}}{d\sqrt{s}}\nonumber\\
   &=& \frac{s(s-4m^2) K_1(\sqrt{s^*})}{2^4T^2m^{3}K_3(2m^*)}
\end{eqnarray}
with $s^*=s/T^2$, and $N_{Q\bar{Q}}^{\textrm{th}}$\ is the collision number between thermal $Q$\ and $\bar{Q}$\ within a certain time.  It shows that the distribution of $\sqrt{s}$ at $t=0.5$ fm$/c$, which is comparable to the mean free time for the heavy quarks, can be roughly described by a thermal distribution at a lower temperature $T=0.22$\ GeV than that of the medium. This is because the heavy quarks carry very little momenta after reversing their direction of motion at this time.  At $t=1.0$ fm$/c$, the distribution of $\sqrt{s}$ can well be described by $T=0.25$ GeV since the heavy quarks are partially thermalized. At an even later time $t=5$ fm$/c$, the distribution of $\sqrt{s}$  approaches the equilibrium one and can thus be described by an effective temperature of $T=0.29$ GeV  close to that of the medium. The approaching of the $\sqrt{s}$ distribution to the thermal distribution from a lower temperature is very different from that of the $p_z$ distribution, which approaches the thermal distribution gradually from an initial hard distribution as shown in Fig.~\ref{fig_pzdis}\footnote{The periodic condition there has no effect in the momentum space.}. 
\par
Similarly,  the distribution of the total momentum $|{\bf P}|$\ of colliding charm pairs, which can also be approximately described by a thermal distribution,  is found to have an effective temperature  of $0.36, 0.33$\ and $0.30$ GeV at time $t=0.5, 1.0$, and $5.0$\ fm$/c$, respectively. The thermal  like distributions of $\sqrt{s}$\ and $|{\bf P}|$\ implies that the regeneration of heavy quarkonia from a medium is always dominated by heavy quarks with low momentum when they are rare. For example, neither the high $p_T$\ $J/\psi$\ at SPS nor the high $p_T$\ $\Upsilon$ at RHIC are expected to be produced from regeneration although the heavy quarks may not be thermalized.

\section{Conclusions}

Based on the Boltzmann equation, we have studied the effective volume $V_{\textrm{eff}}$ of a correlated classical heavy quark pair in a hot medium on their collision rate for rare heavy quark events, which is more realistic than simply considering the volume of the fierball. The $V_{\textrm{eff}}$ is finite due to their initial spatial and momentum correlations even though the system is an open one like in heavy ion collisions.  We have found that  $V_{\textrm{eff}}$ is proportional to the time $t$ when it is much shorter than their mean free time $\tau$ between collisions with the medium partons, i.e., $t\ll \tau$. The increase becomes slower for $t\sim\tau$, and eventually $V_{\textrm{eff}}$ increases linearly with $t^{3/2}$ for $t\gg \tau$. Consequently, the chance for a heavy quark pair to collide with each other per unit time increases monotonously with time $t$. Also, the chance for the heavy quarks to collide again depends sensitively on their initial momentum and the temperature of the medium. Heavy quarks of lower initial momentum in a medium of higher temperature have a larger chance to collide.  Furthermore, the distribution of heavy quark pair center of mass energy corresponds to an effective temperature which is lower than the actual temperature of the medium. All these properties are important for quarkonium regeneration in collisions where heavy quarks are rarely produced. The present study was based on heavy quark scattering cross sections with partons and among themselves that are not from specific model calculations. Although the above results are not expected to qualitatively change, a quantitative study requires more accurate cross sections, which we leave as a future work.

\section*{Acknowledgements}

We thank Ralf Rapp and Nu Xu for helpful discussions. This work was supported by the U.S. National Science Foundation under Grant No. PHY-1068572, the US Department of Energy under Contract No. DE-FG02-10ER41682, and the Welch Foundation under Grant No. A-1358.

\appendix

\section{ Conditions for heavy quark collisions}
\label{ap_b}

Consider one pair of free heavy quark $Q$ and antiquark $\bar{Q}$ with their 4-dimensional coordinates $x_{Q0}$ and $x_{\bar{Q}0}$, and velocities $u_Q$ and $u_{\bar{Q}}$, respectively. Their trajectories in space are
\begin{eqnarray}
   {\bf x}_{Q}(t_Q)&=& {\bf x}_{Q0}+\frac{ {\bf u}_Q }{u_Q^0}(t_Q-t_{Q0}),\nonumber\\
   {\bf x}_{\bar{Q}}(t_{\bar{Q}})&=& {\bf x}_{\bar{Q}0}+\frac{ {\bf u}_{\bar{Q}} }{u_{\bar{Q}}^0}(t_{\bar{Q}}-t_{\bar{Q}0}).
\end{eqnarray}
Then
\begin{eqnarray}
   s^2(t_Q, t_{\bar{Q}})&\equiv& (x_Q(t_Q)-x_{\bar{Q}}(t_{\bar{Q}}))^2\nonumber\\
   &=& (t_Q-t_{\bar{Q}})^2-[{\bf x}_Q(t_Q)-{\bf x}_{\bar{Q}}(t_{\bar{Q}})]^2.\nonumber\\
\end{eqnarray}
When the minimum distance $b$ between  $Q$ and  $\bar{Q}$ in their center of mass frame is reached, it is a saddle point of $s^2(t_Q, t_{\bar{Q}})$.  Requiring
\begin{eqnarray}
   \partial_{t_Q}s^2=\partial_{t_{\bar{Q}}}s^2=0,
\end{eqnarray}
we find the minimum distance $b$ and the corresponding times $t_Q$ and $t_{\bar{Q}}$ to be
\begin{eqnarray}
   b &=& \left(-(\Delta x)^2 -\left[(\Delta x\cdot u_Q)^2+(\Delta x\cdot u_{\bar{Q}})^2\right.\right.\nonumber\\
   &&\left.\left.-\phantom{^2}2(\Delta x\cdot u_Q)(\Delta x\cdot u_{\bar{Q}})u_{Q\bar{Q}}\right]/\left({u_{Q\bar{Q}}^2-1}\right)\right)^{1/2},\nonumber\\
   t_Q&=& t_{Q0}+\frac{-\Delta x\cdot u_Q + ( \Delta x\cdot u_{\bar{Q}})u_{Q\bar{Q}}}{u_{Q\bar{Q}}^2-1}u_Q^0,\nonumber\\
   t_{\bar{Q}}&=& t_{\bar{Q}0}+\frac{\Delta x\cdot u_{\bar{Q}} - ( \Delta x\cdot u_Q)u_{Q\bar{Q}}}{u_{Q\bar{Q}}^2-1}u_{\bar{Q}}^0,
\end{eqnarray}
with $\Delta x=x_{\bar{Q}0}-x_{Q0}$\ and $u_{Q\bar{Q}}=u_Q\cdot u_{\bar{Q}}$. During the time interval $(t, t+\Delta t)$, the two particles are regarded as undergoing a collision if and only if $b\le \sqrt{\sigma_{Q\bar{Q}}/\pi}$ and $t<(t_Q+t_{\bar{Q}})/2 <t+\Delta t$\ are satisfied.

\section{Proof of the linear behavior of the effective volume $V_{\textrm{eff}}$\ at short times}
\label{ap_veff}

 To investigate the time dependence of $V_{\textrm{eff}}$, we consider a pair of heavy quarks in a medium and follow their motions. When the time $t$ is much smaller than the mean free time $\tau$, a heavy quark can have at most one collision with the partons in the medium. Therefore, if we slow down the time by a factor $\lambda$ to $t^\prime=\lambda t$ and also stretch the coordinates by $\lambda$ to $l^\prime=\lambda l$, so that the velocities of the particles remain unchanged, then the number of heavy quark collisions in the original and the scaled space-time are related by 
\begin{eqnarray}
   &&\Delta N_{Q\bar{Q}}'(t', \Delta t', \sigma_{Q\bar{Q}}', \sigma_{gQ}', f_{g}')\nonumber\\
   &=&\Delta N_{Q\bar{Q}}(t,\Delta t, \sigma_{Q\bar{Q}}, \sigma_{gQ}, f_g)
\label{eq_video}
\end{eqnarray}
with $\Delta t'=\lambda\Delta t$, $\sigma'_{Q\bar{Q}}=\lambda^2\sigma_{Q\bar{Q}}$, $\sigma'_{gQ}=\lambda^2\sigma_{gQ}$, and $f'_g=\lambda^{-3}f_g$. On the other hand, since  the number of heavy quark collisions at a given time is, up to a constant, given by
\begin{eqnarray}
   \Delta N_{Q\bar{Q}}\propto \Delta t \sigma_{Q\bar{Q}}(\sigma_{gQ}f_g)^2.
\end{eqnarray}
 we have
\begin{eqnarray}
   &&\Delta N_{Q\bar{Q}}'(t', \Delta t', \sigma_{Q\bar{Q}}', \sigma_{gQ}', f_g')\nonumber\\
   &=& \frac{\Delta t' \sigma_{Q\bar{Q}}'(\sigma'_{gQ}f_g')^2}{\Delta t \sigma_{Q\bar{Q}}(\sigma_{gQ}f_g)^2}\Delta N_{Q\bar{Q}}(t', \Delta t, \sigma_{Q\bar{Q}}, \sigma_{gQ}, f_g)\nonumber\\
   &=& \lambda \Delta N_{Q\bar{Q}}(\lambda t, \Delta t, \sigma_{Q\bar{Q}}, \sigma_{gQ}, f_g).
   \label{eq_scaling}
\end{eqnarray}
Combining Eqs.~(\ref{eq_video}) and (\ref{eq_scaling}), we obtain
\begin{eqnarray}
   &&\Delta N_{Q\bar{Q}}(\lambda t, \Delta t, \sigma_{Q\bar{Q}}, \sigma_{gQ},f_g)\nonumber\\
   &=& \lambda^{-1} \Delta N_{Q\bar{Q}}(t, \Delta t, \sigma_{Q\bar{Q}}, \sigma_{gQ}, f_g),
\end{eqnarray}
and therefore 
\begin{eqnarray}
   V_{\textrm{eff}}(\lambda t) &=& \lambda V_{\textrm{eff}}(t).
\end{eqnarray}
This proves our claim that the effective volume $V_{\textrm{eff}}$\ is linearly proportional to  $t$, as long as $t$\ is much smaller than the mean free time $\tau$  between the collisions of heavy quarks with medium partons. Because the scaling in Eq.(\ref{eq_video})  does not change either the velocities of particles or the angular distribution after their collisions, this proof is independent of the details of the cross sections $\sigma_{gQ}$\ or $\sigma_{Q\bar{Q}}$.

\bibliographystyle{elsarticle-num.bst}
\bibliography{ref}
\end{document}